\documentclass[preprint]{aastex}

\usepackage{emulateapj5}
\usepackage{apjfonts}

\newcommand{\acena}{\mbox{$\alpha$~Cen~A}}
\newcommand{\acenb}{\mbox{$\alpha$~Cen~B}}

\newcommand{\bhyi}{\mbox{$\beta$~Hyi}}
\newcommand{\cms}{\mbox{cm\,s$^{-1}$}}

\newcommand{\ms}{\mbox{m\,s$^{-1}$}}
\newcommand{\muHz}{\mbox{$\mu$Hz}}

\let\epsilon\varepsilon

\slugcomment{Accepted by ApJ Letters}

\shorttitle{Velocity oscillations in $\alpha$~Cen~A}
\shortauthors{Butler et al.}

\begin{document}

\title{Ultra-high-precision velocity measurements of oscillations in
$\alpha$ Cen A}

\author{
R. Paul Butler,\altaffilmark{1}
Timothy R. Bedding,\altaffilmark{2}
Hans Kjeldsen,\altaffilmark{3}
Chris McCarthy,\altaffilmark{1}
Simon J. O'Toole,\altaffilmark{2}
Christopher G. Tinney,\altaffilmark{4}
Geoffrey W. Marcy,\altaffilmark{5}
Jason T. Wright\altaffilmark{5}
}

\altaffiltext{1}{Carnegie Institution of Washington,
Department of Terrestrial Magnetism, 5241 Broad Branch Road NW, Washington,
DC 20015-1305; paul@dtm.ciw.edu, chris@dtm.ciw.edu}

\altaffiltext{2}{School of Physics A28, University of Sydney, NSW 2006,
Australia; bedding@physics.usyd.edu.au, otoole@sternwarte.uni-erlangen.de}

\altaffiltext{3}{Theoretical Astrophysics Center, University of Aarhus,
DK-8000 Aarhus C, Denmark; hans@phys.au.dk}

\altaffiltext{4}{Anglo-Australian Observatory, P.O.\,Box 296, Epping, NSW
1710, Australia; cgt@aaoepp.aao.gov.au}

\altaffiltext{5}{Department of Astronomy, University of California,
Berkeley, CA 94720; and Department of Physics and Astronomy, San Francisco,
CA 94132; gmarcy@astron.berkeley.edu, jtwright@astron.berkeley.edu}

\begin{abstract} 
We have made differential radial velocity measurements of the star
$\alpha$~Cen~A using two spectrographs, UVES and UCLES, both with iodine
absorption cells for wavelength referencing.  Stellar oscillations are
clearly visible in the time series.  After removing jumps and slow trends
in the data, we show that the precision of the velocity measurements per
minute of observing time is 0.42\,\ms{} for UVES and 1.0\,\ms{} for UCLES,
while the noise level in the Fourier spectrum of the combined data is
1.9\,\cms.  As such, these observations represent the most precise
velocities ever measured on any star apart from the Sun.
\end{abstract}

\keywords{stars: individual ($\alpha$~Cen~A) --- stars: oscillations---
techniques: radial velocities}

\section{Introduction}

The search for extra-solar planets has driven tremendous improvements in
high-precision measurements of stellar differential radial velocities
\citep[e.g.,][]{M+B2000}.  In parallel, the same techniques have been used
with great success to measure stellar oscillations.  Recent achievements,
reviewed by \citet{B+C2003} and \citet{B+K2003}, include observations of
oscillations in Procyon with ELODIE \citep[][see also
\citealp{BMM99}]{MSL99}; \bhyi{} with UCLES \citep{BBK2001} and CORALIE
\citep{CBK2001}; and \acena{} and~B with CORALIE
\citep{B+C2001,B+C2002,C+B2003}.  Here we report observations of \acena{}
made with UVES and UCLES which represent the most precise differential
radial velocities ever measured on any star apart from the Sun.

\section{Observations}

We observed \acena{} in May 2001 from two sites.  At the European Southern
Observatory in Chile we used UVES (UV-Visual Echelle Spectrograph) at the
8.2-m Unit Telescope~2 of the Very Large Telescope (VLT)\footnote{Based on
observations collected at the European Southern Observatory, Paranal, Chile
(ESO Programme 67.D-0133)}\@.  At Siding Spring Observatory in Australia we
used UCLES (University College London Echelle Spectrograph) at the 3.9-m
Anglo-Australian Telescope (AAT).  In both cases, an iodine cell was used
to provide a stable wavelength reference \citep{BMW96}.

\begin{figure*}
\epsscale{0.8}
\plotone{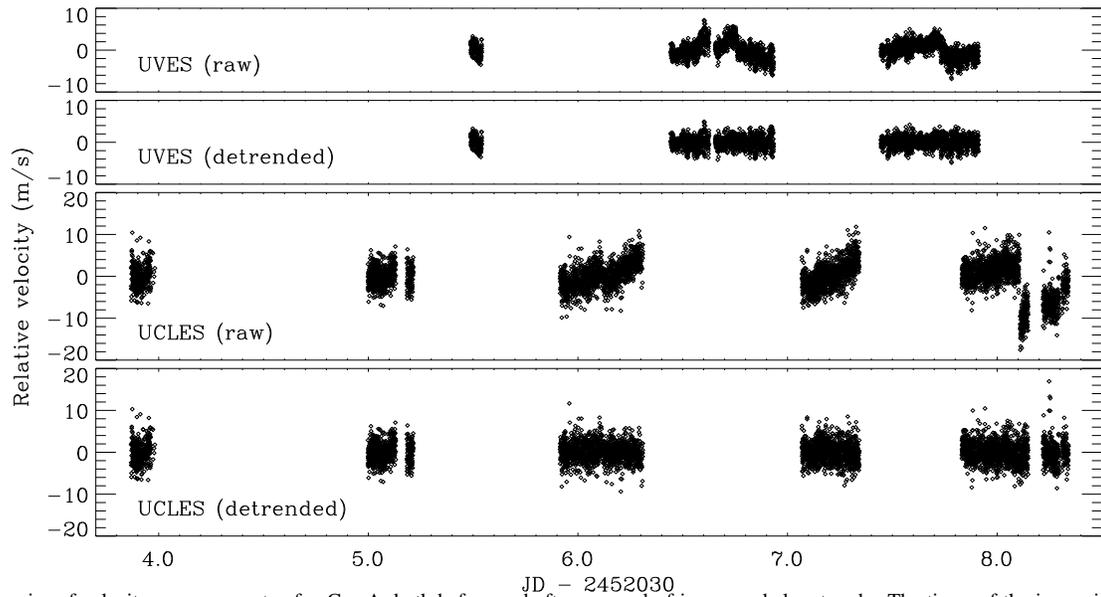}
\caption[]{\label{fig.series} Time series of velocity measurements of
\acena, both before and after removal of jumps and slow trends.  The times
of the jumps in UVES are at 6.655 and 6.760, and those for UCLES are at
6.122 and 8.110.}
\end{figure*}

At the VLT we obtained 3013 spectra of \acena, with typical exposure times
of 1--3\,s and a median cadence of one exposure every 26\,s.  At the AAT we
obtained 5169 spectra of \acena, with typical exposure times of 6\,s and a
median cadence of one exposure every 20\,s.  The resulting velocities are
shown in Fig.~\ref{fig.series}, and the effects of bad weather can be seen
(we were allocated four nights with the VLT and six with the AAT).


The UCLES velocities show upward trends during most nights, which we
believe to be related to the slow movement of the CCD dewar as liquid
nitrogen boiled off.  Nights three and (particularly) five also show a jump
that coincides with the refilling of the CCD dewar in the middle of the
night.  These UCLES observations differed from our earlier run on
$\beta$~Hyi \citep{BBK2001}, and also from standard UCLES planet-search
observations, in that the CCD was rotated by 90 degrees to speed up readout
time.  This had the side-effect of causing the CCD readout to be in the
same direction as the dispersion, and also of making this direction
vertical (so that flexure in the dewar due to changes in its weight would
have shifted the spectrum in the dispersion direction).  It seems there
was an effect on the velocities that was not completely corrected by the
iodine reference method, which suggests that either the PSF
description or the spectrum extraction from the CCD images was inadequate.
We note that improvements to the PSF description are an active area of work
to enhance immunity to spectrometer changes.  The UVES data, meanwhile,
show slow trends and two smaller jumps which are presumably also
instrumental -- at least one of the jumps can be identified with a
correction to the position of the star on the slit.  While these jumps
and slow trends would seriously compromise a planet search, they
fortunately have neglible effect on our measurement of oscillations.

To remove the slow trends, we have subtracted a smoothed version of the
data, and this was done separately on either side of the jumps so that
they, too, were removed.  The detrended time series are shown in
Fig.~\ref{fig.series}.  We have verified by calculating power spectra that
this process of high-pass filtering effectively removes all power below
about 0.2 mHz.  It also removes any power at higher frequencies that arises
from the jumps, which would otherwise degrade the oscillation spectrum.

\section{Analysis and discussion}	\label{sec.time-series}

\begin{figure*}
\epsscale{1} 
\plotone{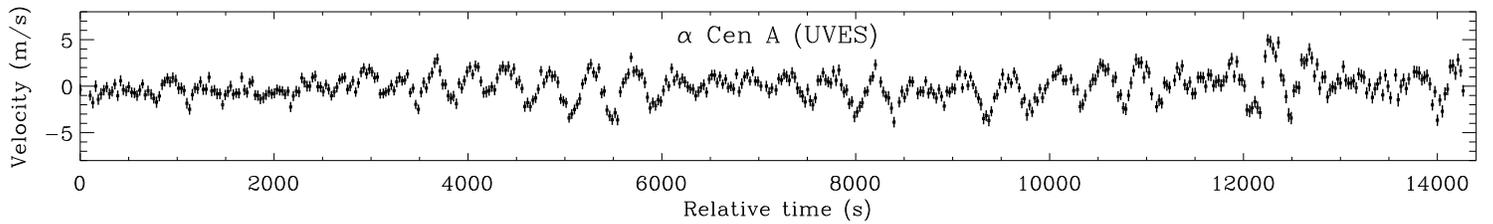}
\caption[]{\label{fig.best}
A four-hour segment of the detrended UVES velocity time series, showing
1-$\sigma$ error bars.  }
\end{figure*}

The time series of velocity measurements clearly show oscillations and the
effects of beating between modes (Fig.~\ref{fig.best}).  The data
presented here have unprecedented precision and we are interested in
obtaining the lowest possible noise in the power spectrum so as to measure
as many modes as possible.  We also wish to estimate the actual precision
of the Doppler measurements.

The algorithm used to extract the Doppler velocity from each spectrum also
provides us with an estimate $\sigma_i$ of the uncertainty in this velocity
measurement.  These are derived from the scatter of velocities measured
from many small ($\sim1.7\,$\AA) segments of the echelle spectrum.  In the
past, we have used these values to generate weights ($w_i = 1/\sigma_i^2$)
for the Fourier analysis \citep{BBK2001,KBB2003}.  In the present analysis,
we take the opportunity to verify that these $\sigma_i$ do indeed reflect
the actual noise properties of the velocity measurements.

Our first step in this process was to measure the noise in the power
spectrum at high frequencies, well beyond the stellar signal.  It soon
became clear that this procedure implied a measurement precision
significantly worse than is indicated by the point-to-point scatter in the
time series itself.  The implication is that some fraction of the velocity
measurements are `bad,' contributing a disproportionate amount of power in
Fourier space.

Since the oscillation signal is the dominant cause of variations in the
velocity series, we need to remove this signal in order to estimate the
noise and to locate the bad points.  We chose to remove the signal
iteratively by finding the strongest peak in the power spectrum and
subtracting the corresponding sinusoid from the time series.  This
procedure was carried out for the strongest peaks in the oscillation
spectrum in the frequency range 0--3.5\,mHz, until spectral leakage into
high frequencies from the remaining power was negligible.  We were left
with a time series of residual velocities, $r_i$, which reflects the noise
properties of the measurements.

\begin{figure*}
\epsscale{0.8} 
\plottwo{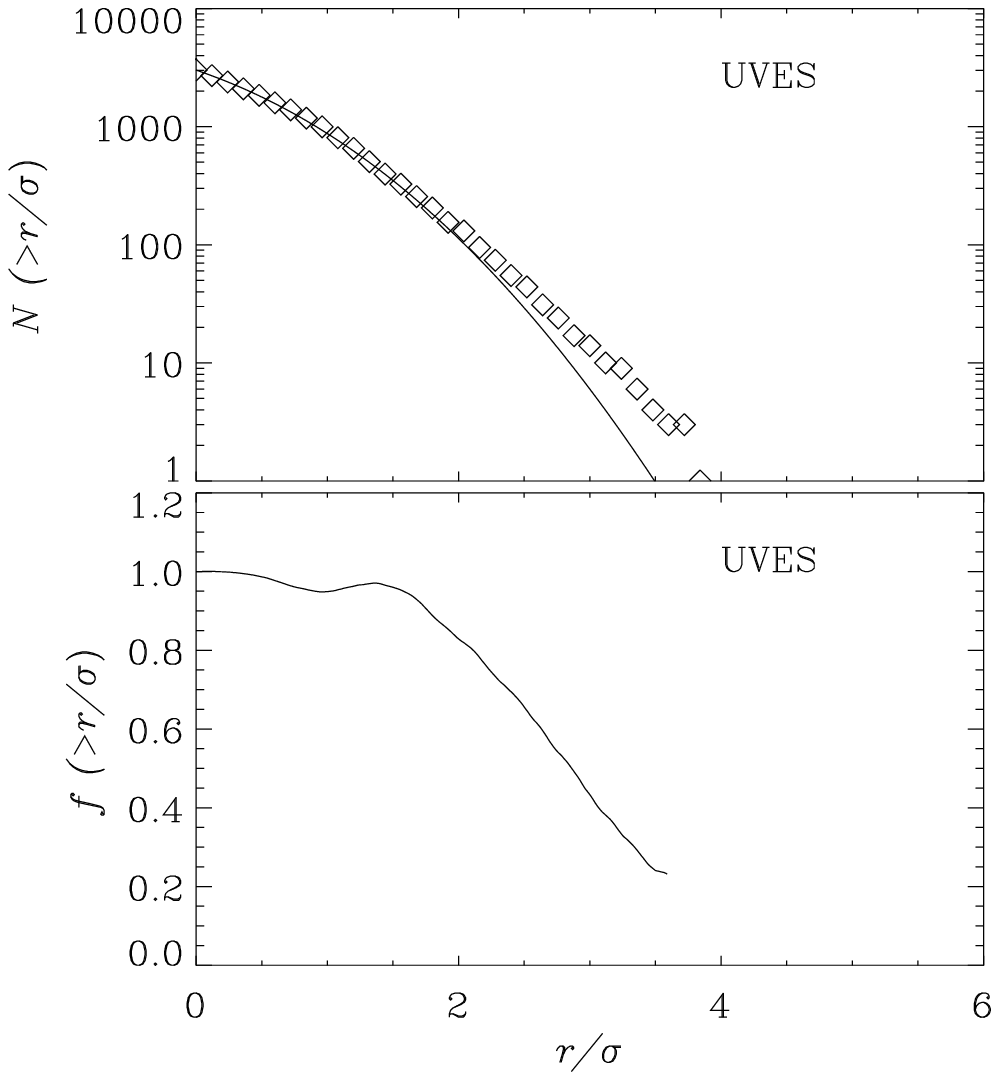}{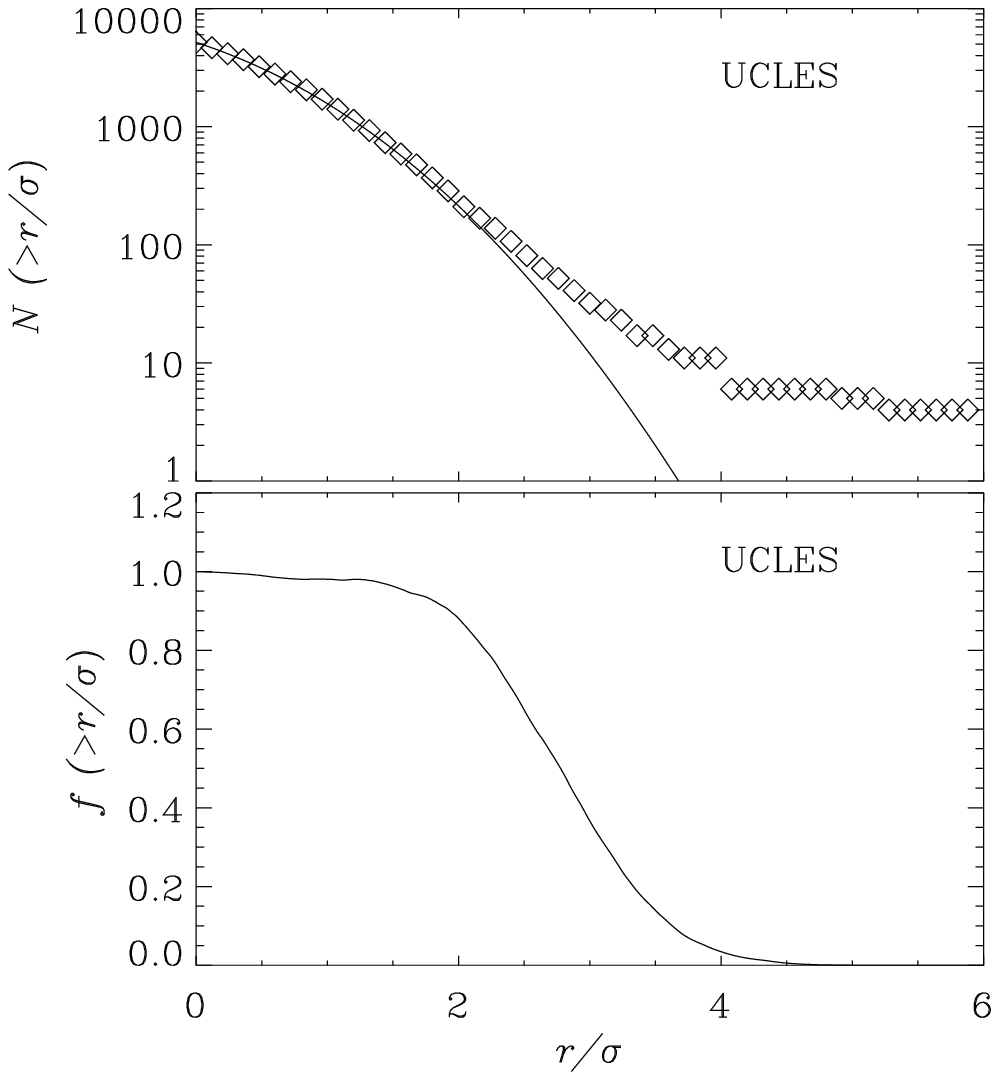}
\caption[]{\label{fig.cumhist}
Upper: cumulative histograms of $|r_i/\sigma_i|$ for UVES (left) and
UCLES) right.  The points show the observations and the solid curve shows
the result expected for Gaussian-distributed noise.  Lower: the ratio
between observed and expected histograms, indicating the fraction of
``good'' data points.}
\end{figure*}

The next step was to analyse the residuals for evidence of bad points,
which we would recognize as those values deviating from zero by more than
would be expected from their uncertainties.  In other words, we examined
the ratio $r_i/\sigma_i$, which we expect to be Gaussian distributed, so
that outliers correspond to suspect data points.  We found that the best
way to investigate this was via the cumulative histogram of
$|r_i/\sigma_i|$, which is shown for both the UVES and UCLES as the points
in the upper panels of Fig.~\ref{fig.cumhist}.  The solid curves in these
figures show the cumulative histograms for the best-fitting Gaussian
distributions.  We indeed see a significant excess of outliers for
$|r_i/\sigma_i|$ beyond about 2, in both data sets.  The lower panels show
the ratio between the points and the curve, which is the fraction $f$ of
data points that could be considered as good.  This fraction is essentially
unity out to $|r_i/\sigma_i| \simeq 2$, and then falls off, indicating that
about half the data points with $|r_i/\sigma_i| > 3$ are bad.

At this point, we could simply make the decision to ignore all data points
with $|r_i/\sigma_i|$ above a certain value, such as 3.0, on the basis that
many of them would be bad points that would increase the noise in the
oscillation spectrum.  We instead chose a more elegant approach, which gave
essentially the same results, in which we used the information in
Fig.~\ref{fig.cumhist} to adjust the weights: those points with large
values of $|r_i/\sigma_i|$ were decreased in weight, with every $w_i$
multiplied by the factor~$f$.  Given that the weights are calculated as
$w_i = 1/\sigma_i^2$, the adjustment was achieved by dividing each
measurement error ($\sigma_i$) by the square root of~$f$.

With these adjustments to the measurement uncertainties, which effectively
down-weight the bad data points, we now expect the noise floor at high
frequencies in the amplitude spectrum of the residuals ($r_i$) to be
substantially reduced.  We measured the average noise~$\sigma_{\rm amp}$ in
the range 7.5--15\,mHz to be 2.11\,\cms{} for UVES and 4.37\,\cms{} for
UCLES\@.  The corresponding values before down-weighting the bad data
points were 2.33\,\cms{} for UVES and 4.99\,\cms{} for UCLES\@.  We can
therefore see that adjusting the weights has lowered the noise by about
10\%.

\begin{figure*}
\epsscale{0.8} 
\plotone{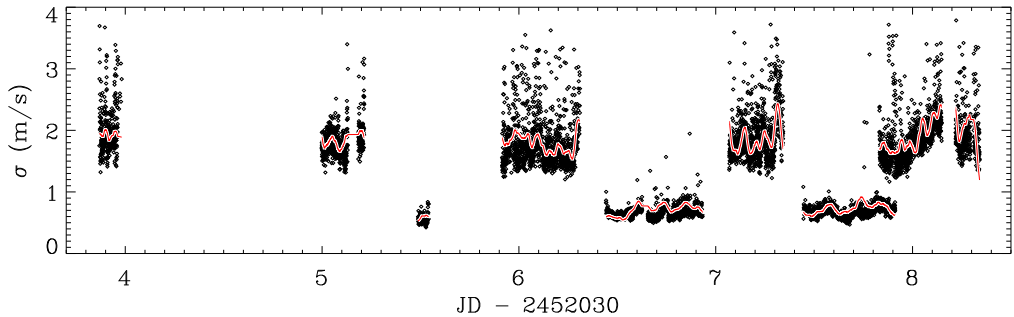}
\caption[]{\label{fig.errors}
Points show measurement errors in each time series, while the (red) curves show
the smoothed residuals (see text) }
\end{figure*}

The final stage in this processing involved checking the calibration of the
uncertainties.  By this, we mean that the estimates $\sigma_i$ should be
consistent with the noise level determined from the amplitude spectrum.  On
the one hand, the mean variance of the data can be calculated as a weighted
mean of the $\sigma_i$, as follows:
\begin{equation}
  \sum_{i=1}^{N} \sigma_i^2 w_i \left/   \sum_{i=1}^{N} w_i \right.,
\end{equation}
where the weights are given by $w_i = 1/\sigma_i^2$ (which means the
numerator is simply equal to $N$).  On the other hand, the variance deduced
from the noise level $\sigma_{\rm amp}$ in the amplitude spectrum is
\citep[Appendix A.1 of][]{K+B95}:
\begin{equation}
  \sigma_{\rm amp}^2 N/\pi.
\end{equation}
We require these to be equal, which gives the condition
\begin{equation}
      \sigma_{\rm amp}^2 \sum_{i=1}^{N} \sigma_i^{-2}  = \pi. 
        \label{eq.condition}
\end{equation}
Using the values of $\sigma_{\rm amp}$ given above, we concluded that
Eq.~(\ref{eq.condition}) would be satisfied for each data set provided the
uncertainties $\sigma_i$ were multiplied by 0.78 for UVES and 0.87 for
UCLES\@.  It is these calibrated $\sigma_i$ that are shown in
Figs.~\ref{fig.best} and~\ref{fig.errors}, and they represent our best
estimate of the high-frequency precision of the data.  Also shown in
Fig.~\ref{fig.errors} as solid lines are the residuals $r_i$ after
smoothing with a running box-car mean.  We can see that there is generally
very good agreement between these two independent measures of the velocity
precision, giving us confidence that we have correctly estimated $\sigma_i$
both in relative terms and in the absolute calibration.

\begin{figure*}
\epsscale{0.35} 
\plotone{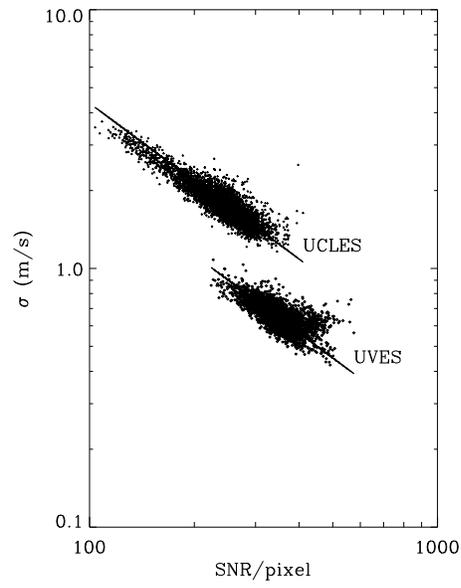}
\caption[]{\label{fig.snr}Measurement errors as a function of SNR\@.  The
two straight lines both have slopes of $-1$, and are displaced from each
other by a factor 1.93 (see text).}
\end{figure*}

We have also investigated the dependence of the velocity precision on the
photon flux in the stellar spectrum.  Figure~\ref{fig.snr} shows $\sigma_i$
versus the signal-to-noise ratio (SNR), where the latter is the square root
of the median number of photons per pixel in the iodine region of the
spectrum.  The points for both UVES and UCLES agree well with a slope of
$-1$ in the logarithm, as expected for Poisson statistics.  We also note
that the offset between the two sets of points arises because of the
difference in dispersion of the spectrographs.  UVES has higher dispersion,
by a factor of 1.55, which means that not only is velocity precision
expected to be better by this factor, but also that the SNR per unit
wavelength is lower (by the square root of this factor).  Combining these
effects, we would expect the two distributions to be in the ratio
$1.55^{1.5}=1.93$, which is exactly what we find in Fig.~\ref{fig.snr}.  In
other words, the lower precision in the UVES data at a given SNR per pixel
results from the higher dispersion but, allowing for this, the intrinsic
precision is the same for the two systems.  In addition, the CCD on the
UVES system was able to record more photons per exposure than UCLES (the
median SNR is considerably higher).

\begin{figure*}
\epsscale{0.8} 
\plotone{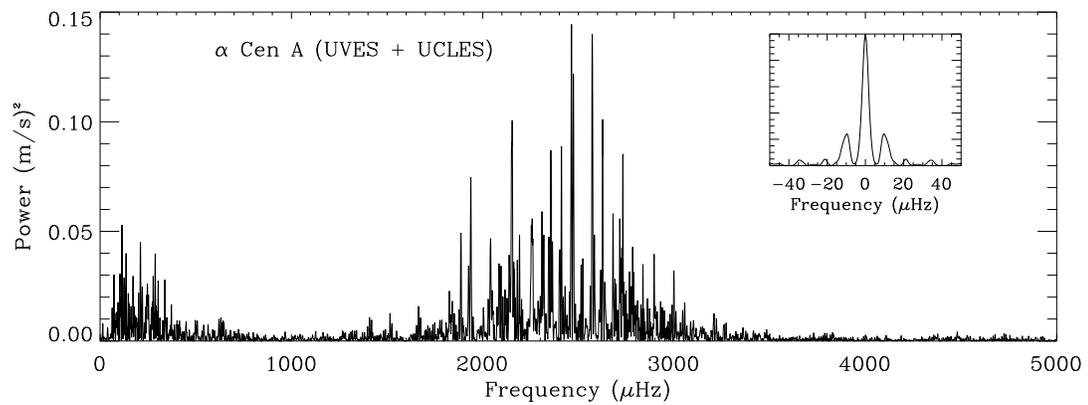}
\caption[]{\label{fig.power}
Power spectrum of the combined velocity time series.  The inset shows the
spectral window, with the frequency scale expanded by a factor of ten. }
\end{figure*}

The result of the processing described above was a time series for both
UVES and UCLES, each of which consisted of the time stamps, the measured
velocities (after correction for jumps and drifts) and the uncertainty
estimates (after the adjustments described).  These two time series could
then be merged in order to produce the oscillation power spectrum, and this
is shown in Fig.~\ref{fig.power}.  The average noise in the amplitude
spectrum in the range 7.5--15\,mHz is 2.03\,\cms.  As discussed at the
beginning of this section, some of this power comes from spectral leakage
of the oscillation signal itself.  Therefore, a more accurate measure of
the noise is obtained from the power spectrum of the residuals,~$r_i$, when
using the final weights, $1/\sigma_i^2$.  The result gives an average noise
in the range 7.5--15\,mHz of 1.91\,\cms.  For comparison, the noise level
reported from CORALIE observations of \acena{} was 4.3\,\cms{}
\citep{B+C2002}, while observations of \acenb{} gave 3.75\,\cms{}
\citep{C+B2003}.

We can also calculate the precision per minute of observing time, which is
shown in Table~\ref{tab.noise}.  For comparison, we include velocity
precision from other oscillation measurements.  The list is not meant to be
exhaustive, but most of the instruments that have been used in recent years
are represented.  The precision depends, of course, on several factors such
as the telescope aperture, target brightness, observing duty cycle,
spectrograph stability and method of wavelength calibration.  It is clear
that the observations reported here, particularly those with UVES, are
significantly more precise than any previous measurements of stars other
than the Sun.  Of course, we refer here to the precision at
frequencies above $\sim0.8$\,mHz, which is the regime of interest for
oscillations in solar-type stars.

The referee has questioned whether adjusting the weights using the method
described above might have affected the accuracy with which the oscillation
frequencies can be measured.  To test this, we have extracted the ten
highest peaks from both the power spectrum in Fig.~\ref{fig.power} and from
the power spectrum obtained without adjusting the weights.  The frequencies
of all ten peaks agreed very well, with a mean difference of 0.11\,\muHz{}
and a maximum difference of 0.4\,\muHz.  The latter is a factor of ten
smaller than the FWHM of the spectral window and probably well below the
natural linewidth of the modes.  Therefore, there is no reason to think
that the reduction in the noise level obtained by adjusting weights has
come at the expense of reduced accuracy in the measured frequencies.

\section{Conclusion}

We have analysed differential radial velocity measurements of
\acena{} made with UVES at the VLT and UCLES at the AAT\@.  Stellar
oscillations are clearly visible in the time series.  Slow drifts and
sudden jumps of a few metres per second, presumably instrumental, were
removed from each time series using a high-pass filter.  We then used the
measurement uncertainties as weights in calculating the power spectrum, but
we found it necessary to modify some of the weights to account for a small
fraction of bad data points.  In the end, we reached a noise floor of
1.9\,\cms{} in the amplitude spectrum and in a future paper we will present
a full analysis of the oscillation frequencies and a comparison with
stellar models.

\acknowledgments

We thank Bill Chaplin for providing a times series that allowed us to
estimate the precision of BiSON observations.  This work was supported
financially by the Australian Research Council, by the Danish Natural
Science Research Council and by the Danish National Research Foundation
through its establishment of the Theoretical Astrophysics Center.  We
further acknowledge support by NSF grant AST-9988087 (RPB), and by SUN
Microsystems.

\begin{table*}
\begin{center}
\caption{\label{tab.noise}Noise levels from observations of solar-like oscillations}
\small
\begin{tabular}{llcl}
\hline
\hline
Star & Spectrograph & Precision in & Reference \\
     &           & 1\,min (\ms{})          \\
\hline
\noalign{\smallskip}
Sun        &  BiSON     &  0.2  & data supplied by W. Chaplin      \\
\acena     &  UVES      &  0.42 & this paper      \\
Sun        &  GOLF      &  0.6  &  data from the GOLF Web site      \\
\acena     &  UCLES     &  1.0  & this paper      \\
\acenb     &  CORALIE   &  1.7  & \citet{C+B2003} \\
\acena     &  CORALIE   &  1.7  & \citet{B+C2002} \\
Procyon    &  ELODIE    &  2.5  & \citet{MSL99} \\
Procyon    &  CORALIE   &  2.7  & \citet{CBK2002} \\
\bhyi      &  UCLES     &  3.0  & \citet{BBK2001} \\
\bhyi      &  CORALIE   &  4.2  & \citet{CBK2001} \\
Procyon    &  AFOE      &  4.2  & \citet{Bro2000} \\
Procyon    &  FOE       &  4.7  & \citet{BGN91}   \\
Procyon    &  HIDES     &  5.1  & \citet{KST2003} \\
\noalign{\smallskip}
\hline
\end{tabular}
\end{center}
\end{table*}


\begin{thebibliography}{17}
\expandafter\ifx\csname natexlab\endcsname\relax\def\natexlab#1{#1}\fi
\expandafter\ifx\csname url\endcsname\relax
  \def\url#1{{\tt #1}}\fi

\bibitem[Barban et~al.(1999)Barban, {Michel}, {Martic}, {Schmitt}, {Lebrun},
  {Baglin}, \& {Bertaux}]{BMM99}
Barban, C., {Michel}, E., {Martic}, M., {Schmitt}, J., {Lebrun}, J.~C.,
  {Baglin}, A., \& {Bertaux}, J.~L., 1999, A\&A, 350, 617.

\bibitem[Bedding et~al.(2001)Bedding, Butler, Kjeldsen, Baldry, O'Toole,
  Tinney, Marcy, Kienzle, \& Carrier]{BBK2001}
Bedding, T.~R., Butler, R.~P., Kjeldsen, H., Baldry, I.~K., O'Toole, S.~J.,
  Tinney, C.~G., Marcy, G.~W., Kienzle, F., \& Carrier, F., 2001, ApJ, 549,
  L105.

\bibitem[Bedding \& Kjeldsen(2003)Bedding, \& Kjeldsen]{B+K2003}
Bedding, T.~R., \& Kjeldsen, H., 2003, Proc. Astron. Soc. Aust., 20, 203.

\bibitem[Bouchy \& {Carrier}(2001)Bouchy, \& {Carrier}]{B+C2001}
Bouchy, F., \& {Carrier}, F., 2001, A\&A, 374, L5.

\bibitem[Bouchy \& {Carrier}(2002)Bouchy, \& {Carrier}]{B+C2002}
Bouchy, F., \& {Carrier}, F., 2002, A\&A, 390, 205.

\bibitem[Bouchy \& {Carrier}(2003)Bouchy, \& {Carrier}]{B+C2003}
Bouchy, F., \& {Carrier}, F., 2003, Ap\&SS, 284, 21.

\bibitem[Brown(2000)]{Bro2000}
Brown, T.~M., 2000, In Teixeira, T., \& Bedding, T.~R., editors, {\em The Third
  MONS Workshop: Science Preparation and Target Selection}, page~1. Aarhus:
  Aarhus Universitet.
\newblock available via the ADS.

\bibitem[Brown et~al.(1991)Brown, Gilliland, Noyes, \& Ramsey]{BGN91}
Brown, T.~M., Gilliland, R.~L., Noyes, R.~W., \& Ramsey, L.~W., 1991, ApJ, 368,
  599.

\bibitem[Butler et~al.(1996)Butler, Marcy, Williams, McCarthy, Dosanjh, \&
  Vogt]{BMW96}
Butler, R.~P., Marcy, G.~W., Williams, E., McCarthy, C., Dosanjh, P., \& Vogt,
  S.~S., 1996, PASP, 108, 500.

\bibitem[Carrier et~al.(2001)Carrier, {Bouchy}, {Kienzle}, {Bedding},
  {Kjeldsen}, {Butler}, {Baldry}, {O'Toole}, {Tinney}, \& {Marcy}]{CBK2001}
Carrier, F., {Bouchy}, F., {Kienzle}, F., {Bedding}, T.~R., {Kjeldsen}, H.,
  {Butler}, R.~P., {Baldry}, I.~K., {O'Toole}, S.~J., {Tinney}, C.~G., \&
  {Marcy}, G.~W., 2001, A\&A, 378, 142.

\bibitem[Carrier et~al.(2002)Carrier, {Bouchy}, {Kienzle}, \&
  {Blecha}]{CBK2002}
Carrier, F., {Bouchy}, F., {Kienzle}, F., \& {Blecha}, A., 2002, In Aerts, C.,
  Bedding, T.~R., \& Christensen-Dalsgaard, J., editors, {\em IAU Colloqium
  185: Radial and Nonradial Pulsations as Probes of Stellar Physics}, volume
  259, page 468. ASP Conf. Ser.

\bibitem[Carrier \& {Bourban}(2003)Carrier, \& {Bourban}]{C+B2003}
Carrier, F., \& {Bourban}, G., 2003, A\&A, 406, L23.

\bibitem[{Kambe} et~al.(2003){Kambe}, {Sato}, {Takeda}, {Izumiura}, {Masuda},
  \& {Ando}]{KST2003}
{Kambe}, E., {Sato}, B., {Takeda}, Y., {Izumiura}, H., {Masuda}, S., \& {Ando},
  H., 2003, In Thompson, M.~J., Cunha, M.~S., \& Monteiro, M. J. P. F.~G.,
  editors, {\em Asteroseismology Across the HR Diagram}, page P331. Kluwer.

\bibitem[Kjeldsen \& Bedding(1995)Kjeldsen, \& Bedding]{K+B95}
Kjeldsen, H., \& Bedding, T.~R., 1995, A\&A, 293, 87.

\bibitem[Kjeldsen et~al.(2003)Kjeldsen, Bedding, Baldry, Bruntt, Butler,
  Fischer, Frandsen, Gates, Grundahl, Lang, Marcy, Misch, \& Vogt]{KBB2003}
Kjeldsen, H., Bedding, T.~R., Baldry, I.~K., Bruntt, H., Butler, R.~P.,
  Fischer, D.~A., Frandsen, S., Gates, E.~L., Grundahl, F., Lang, K., Marcy,
  G.~W., Misch, A., \& Vogt, S.~S., 2003, AJ, 126, 1483.

\bibitem[Marcy \& {Butler}(2000)Marcy, \& {Butler}]{M+B2000}
Marcy, G.~W., \& {Butler}, R.~P., 2000, PASP, 112, 137.

\bibitem[Martic et~al.(1999)Martic, Schmitt, Lebrun, Barban, Connes, Bouchy,
  Michel, Baglin, Appourchaux, \& Bertaux]{MSL99}
Martic, M., Schmitt, J., Lebrun, J.-C., Barban, C., Connes, P., Bouchy, F.,
  Michel, E., Baglin, A., Appourchaux, T., \& Bertaux, J.-L., 1999, A\&A, 351,
  993.

\end{thebibliography}
\end{document}